\documentclass[prb,amsmath,amssymb,twocolumn]{revtex4}

\usepackage[pdftex]{graphicx}
\usepackage[pdftex]{epsfig}
\usepackage{epstopdf}
\usepackage{dcolumn}% Align table columns on decimal point
\usepackage{bm}% bold math

\begin{document}

\title{Spin blockade and lifetime-enhanced transport in a few-electron Si/SiGe double quantum dot}

\author{Nakul Shaji$^1$}
\author{C. B. Simmons$^1$}
\author{Madhu Thalakulam$^1$}
\author{Levente J. Klein$^1$}
\author{Hua Qin$^1$}
\author{H. Luo$^1$}
\author{D. E. Savage$^1$}
\author{M. G. Lagally$^1$}
\author{A. J. Rimberg$^2$}
\author{R. Joynt$^1$}
\author{M. Friesen$^1$}
\author{R. H. Blick$^1$}
\author{S. N. Coppersmith$^1$}
\author{M. A. Eriksson$^1$}
\affiliation{$^1$University of Wisconsin-Madison, Madison, Wisconsin 53706, USA}
\affiliation{$^2$Dartmouth College, Hanover, New Hampshire 03755, USA}

\maketitle

\textbf{Spin blockade occurs when an electron is unable to access an energetically favorable path through a quantum dot due to spin conservation, resulting in a blockade of the current through the dot.\cite{r1,r2,r3,r4,r5,r6}  Spin blockade is the basis of a number of recent advances in spintronics, including the measurement and the manipulation of individual electron spins.\cite{r7,r8}  We report measurements of the spin blockade regime in a silicon double quantum dot, revealing a complementary phenomenon: lifetime-enhanced transport.  We argue that our observations arise because the decay times for electron spins in silicon are long, enabling the electron to maintain its spin throughout its transit across the quantum dot and access fast paths that exist in some spin channels but not in others.  Such long spin lifetimes are important for applications such as quantum computation and, more generally, spintronics.}

Semiconductor quantum dots or Ôartificial atomsÕ provide highly tunable structures for trapping and manipulating individual electrons.\cite{r9,r10,r11}  Such quantum dots are promising candidates as qubits for quantum computation,\cite{r12,r13,r14} due in part to the long lifetimes and slow dephasing of electron spins in semiconductors.\cite{r7,r15} Si quantum dots are predicted to have especially long lifetimes and slow dephasing, due to low spin-orbit interaction and low nuclear spin density.\cite{r16,r17}  In the last several years, much activity has focused on the development of quantum dots in Si/SiGe (refs.~\onlinecite{r18,r19,r20,r21,r22}) and recent advances in materials quality and fabrication techniques have enabled the observation of coherent spin phenomena in such quantum dots.\cite{r23}

Spin-to-charge conversion, in which spin states are detected through their effect on charge motion, enables measurement of individual electron spins in quantum dots.\cite{r15}  Spin blockade is the canonical example of spin-to-charge conversion in transport, where charge current is blocked in a double quantum dot by a metastable spin state. The blockade occurs when one electron is confined in the left dot and an additional electron enters the right dot forming a spin triplet state T(1,1) (Fig.~1a). Exiting the dot requires reaching the triplet T(2,0), with both electrons in the left dot, a state which is higher in energy. The electron is thus trapped in the right dot, unless relaxation from T(1,1) to S(1,1) occurs, opening a downhill channel through S(2,0). As we show below, this aspect of spin blockade in Si is virtually identical to that previously observed in other systems.\cite{r1,r2,r3}

The unexpected effect presented here is lifetime-enhanced transport (LET).  The energy level diagram for LET is the same as for spin blockade, except that current flows in the opposite direction (Fig.~1b). Flow through the triplet channel is now energetically downhill, whereas flow through the singlet channel is very slow, because it requires either an uphill transition or tunneling directly from the left dot to the right lead.  Transport current will be observable only if electrons flow almost exclusively through the triplet channel, requiring even slower triplet-singlet relaxation rates than those needed to observe spin blockade.  

The tunable quantum dot used in these experiments was formed in a Si/SiGe heterostructure. The gate structure (Fig. 2a) has the shape often associated with a single quantum dot, and the corresponding Coulomb diamonds are shown in Fig.~2b.  By tuning the gate voltages, the single dot was split into two tunnel-coupled quantum dots. Such transformations of a lateral single quantum dot into multiple quantum dots have been demonstrated in similar systems.\cite{r24,r25}  Here, by changing voltages on gates G and CS, and keeping those on $\text{B}_\text{L}$, T, and $\text{B}_\text{R}$ fixed, the electron occupations are tuned while keeping the tunnel barriers constant (Fig.~2d). The left dot is coupled more strongly to gate G, and the right dot is coupled more strongly to gate CS. The electron occupancies indicated in the figure correspond to an equivalent charge configuration with a single unpaired spin in the (1,0) state.

The region of interest in this letter is indicated by the blue dashed circle in Fig.~2d. This Ôtriple pointÕ corresponds to degeneracy between the (1,0), (1,1) and (2,0) charge states.\cite{r26,r27}  In this regime, an electron with a spin   or   is confined in the left dot, and if the incoming electron has an anti-parallel spin, a spin singlet S:  $(|\uparrow \downarrow\rangle -|\downarrow \uparrow \rangle)/\sqrt{2}$  is formed, while a parallel spin forms any of the spin triplets $\text{T}_+$:  $|\uparrow \uparrow \rangle$, $\text{T}_-$:  $|\downarrow \downarrow \rangle$ or $\text{T}_0$:  $(|\uparrow \downarrow\rangle +|\downarrow \uparrow \rangle)/\sqrt{2}$, which are degenerate at zero magnetic field. The singlet-triplet energy splitting is larger for two electrons occupying the same dot (2,0), than when they are in different dots (1,1), resulting in the energy level schematics shown in Fig~1.

Spin blockade arises because spin is conserved during tunneling, preventing the direct transition from the triplet T(1,1) to the singlet S(2,0).  We observe this blockade as shown in Fig.~3a-c.  These measurements are taken at finite bias, where the triple points expand into bias triangles.\cite{r28} When T(1,1) is loaded and no relaxation occurs from T(1,1) to S(1,1), spin blockade is observed (as marked by the orange triangle), and the bias triangle is truncated as shown schematically in Fig.~3b. The observed current in the blockaded region is limited by the noise floor in the measurement (7~fA r.m.s). Spin blockade is fully lifted when the T(2,0) state is brought below the T(1,1) state (blue star).  

\begin{figure}[t] 
\centering
\includegraphics[width=3.3in,keepaspectratio]{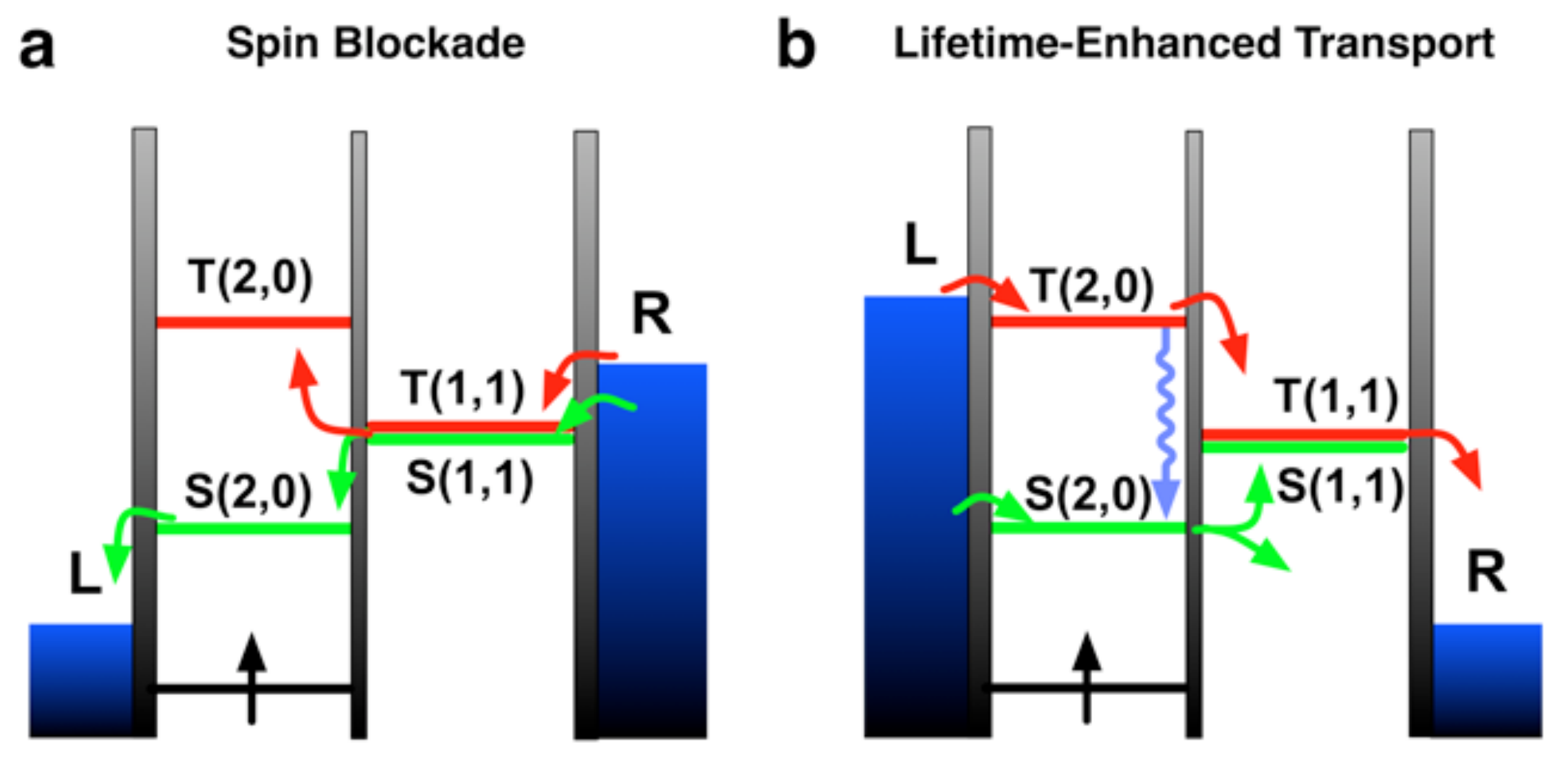}
\caption{
\textbf{Pauli spin blockade regime in a double quantum dot.  a},
Energy level schematic of a double quantum dot with one electron confined in the left dot (black). The applied bias causes electron flow from the right lead (R) to the left (L). Incoming electrons can form either a spin singlet S(1,1) (green) or a spin triplet T(1,1) (red). Electrons forming a singlet have an accessible fast channel through S(2,0) to the left lead.  In contrast, electrons entering the triplet T(1,1) cannot exit through T(2,0), resulting in metastable occupation of the T(1,1) state, and causing spin blockade of the current.
\textbf{b}, The identical energy level configuration as in \textbf{a}, but with the direction of current flow reversed. Electrons entering S(2,0) have no fast path through the system. If S(2,0) loads more rapidly than it unloads, current will be blockaded. In contrast, electrons entering T(2,0) have an accessible fast channel through T(1,1), provided spin relaxation to S(2,0) (blue wavy arrow) does not occur. Electron transport through the triplets, supported by a long spin lifetime, is denoted LET. }
\end{figure}

Now consider the same energy level configuration, but with opposite bias across the dot (Fig.~1b).  In previous work, this configuration has been shown to be blockaded.\cite{r2,r8} In contrast, here in Si we observe a strong ÔtailÕ of current in this configuration, corresponding to the additional parallelograms (green outline) in Fig.~3d,e. As shown in detail below, the condition for observing this tail is that the metastable S(2,0) state must be loaded more slowly than it empties. The relaxation rate from the T(2,0) state into S(2,0) sets a lower bound for this loading rate.  Because the measured current at the point labeled (+) is significant only when the spin lifetime of T(2,0) is long, we denote this tail of current the triplet tail and the effect LET.

The dimensions of the triplet tail in the charge stability diagram (Fig.~3d,e) provide a measurement of the energy difference between the (2,0) triplet and singlet states ($E_\text{ST} = E_\text{T}-E_\text{S}$).  Both the length of the tail and the distance between the tail and the edge of the bias triangle correspond to $E_\text{ST}$ (Fig.~3e). This (2,0) singlet-triplet energy gap as extracted from the data is $240\pm 30$~$\mu$eV. 

\begin{figure}[t] 
\centering
\includegraphics[width=3.4in,keepaspectratio]{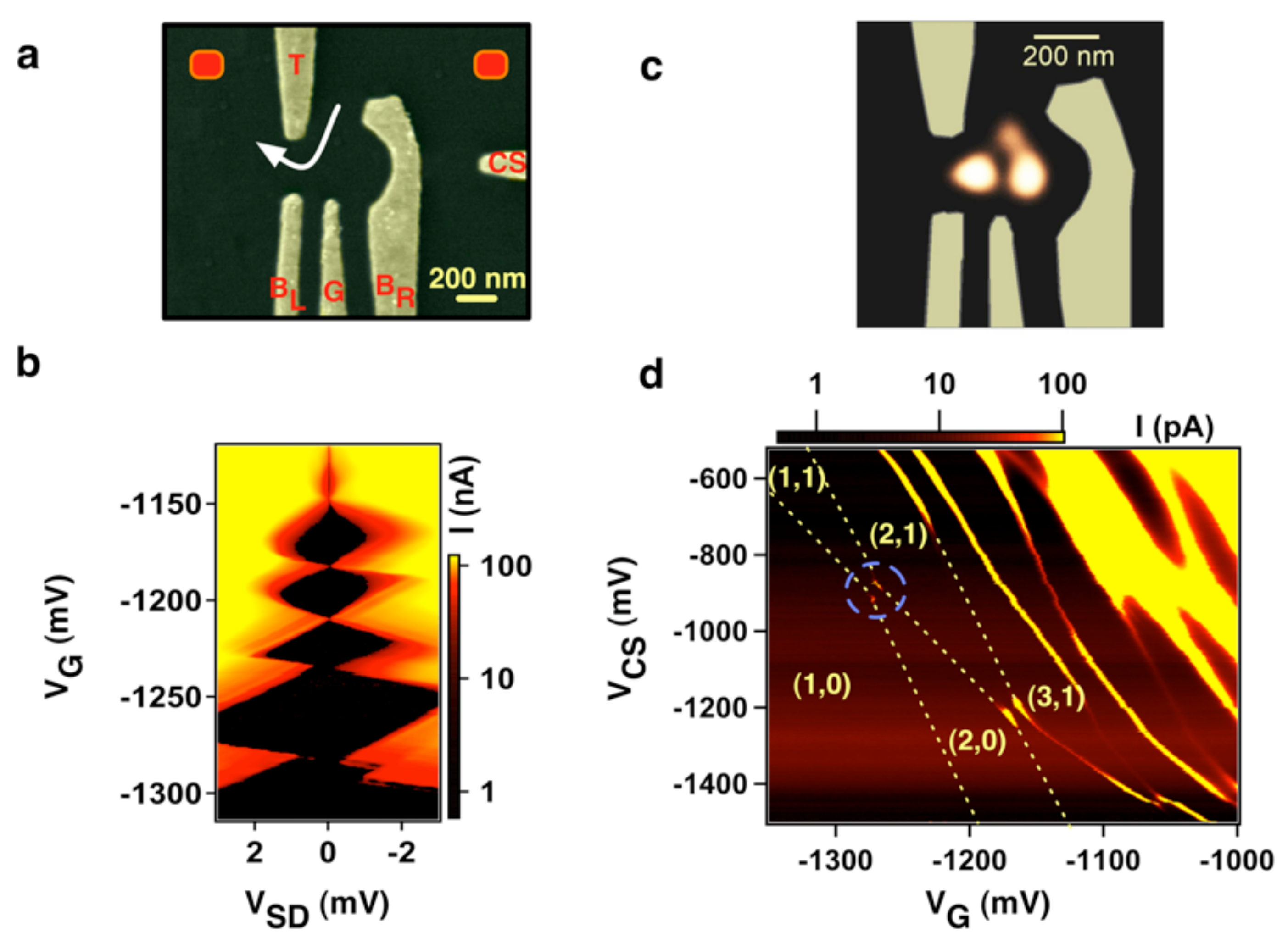}
\caption{
\textbf{Formation of a double quantum dot.  a},
False-color micrograph of a device similar to the one used in the experiments. A two-dimensional electron gas was formed in a 12~nm strained silicon quantum well with a sheet carrier density of $4\times 10^{11}\,\text{cm}^{-2}$ and a mobility of 40,000~cm$^2$V$^{-1}$s$^{-1}$. Ohmic contacts (indicated schematically by red squares) were formed by annealing an alloy of Au:Sb(1\%) at $550^\circ$C. Metal gates used to form the quantum dots were realized by depositing palladium on the sample surface and are labeled on the micrograph.  The white arrow indicates the direction of electron flow when the $V_\text{SD} > 0$.
\textbf{b},  Magnitude of the measured current as a function of source-drain voltage $V_\text{SD}$ and the voltage on gate G. The black regions indicate Coulomb blockade where the number of electrons in the dot is fixed. Outside this blockade, single electron tunneling through the dot occurs. 
\textbf{c},  A numerical simulation of the charge density for the gates as shown and for gate voltages corresponding to the double quantum dot data (\textbf{d}).  Two electrons prefer to occupy opposite sides of the open region between the gates.
\textbf{d}  The single quantum dot was deformed into two tunnel-coupled dots in series by using a combination of negative voltages on gates T, B$_\text{L}$, and B$_\text{R}$. The magnitude of the measured current through the double quantum dot is plotted as a function of the voltages on gates G and CS, with  $V_\text{SD}= 0.1$~mV. The dot coupled more to gate G (CS) is the left (right) dot. As described in the text, the notation ($m,n$) represents the effective left and right dot electron occupancy, and the triple point studied in detail in this letter is indicated by the blue circle. }
\end{figure}

\begin{figure*}[t] 
\centering
\includegraphics[width=6.5in,keepaspectratio]{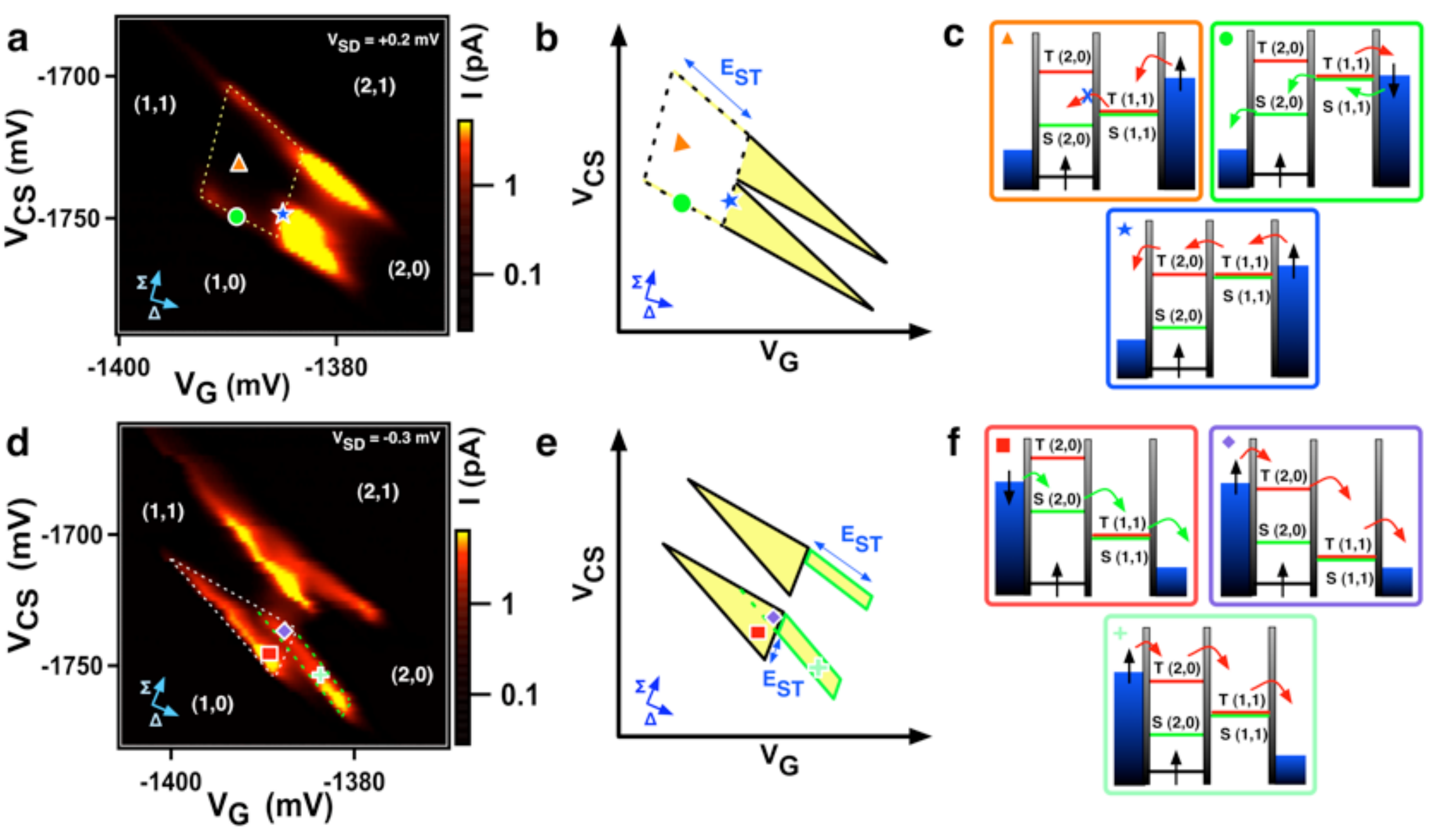}
\caption{
\textbf{Spin blockade and LET.  a},
Positive bias ($V_\text{SD} = 0.2$~mV) triangles representing the (1,0)-(1,1)-(2,0)-(2,1) charge transition. Current through the device is blocked due to Pauli spin blockade (as marked by the orange triangle) in the region outlined by yellow dashed lines. The blockade is lifted when the triplet state becomes accessible (blue star). The lower triangle is referred to as the Ôelectron triangleÕ and the upper triangle is called the Ôhole triangleÕ. The blue arrows represent the energy axis $\Sigma$ where the energy levels of both dots are changed together, and the detuning axis $\Delta$ where dot levels are moved in opposite directions.  Note that the edges of the spin blockade regime (green circle) show measurable current flow due spin exchange with the leads.\cite{r2}
\textbf{b},  Schematic representation of the positive bias triangles. Dashed lines mark the blockaded region while filled yellow regions indicate no blockade. The energy gap between spin singlet and triplet ($E_\text{ST}$) is indicated along the edge of the triangles.
\textbf{c},  Transport details and energy level schematics for the electron triangle corresponding to the points denoted by the green circle, orange triangle, and blue star in \textbf{a} and \textbf{b}.  The $\Sigma$ directions in parts \textbf{a} and \textbf{b} correspond to moving the levels on the left and right up and down together, and the $\Delta$ directions correspond to moving the levels on the left and the right in opposite directions.  The same pattern holds for parts \textbf{d-f}.
\textbf{d},  Negative bias ($V_\text{SD}  = -0.3$~mV) triangles through the same charge transition. Current flow through full electron and hole triangles is observed (white dashed lines are shown on the electron triangle).  In addition, strong ÔtailsÕ of current (green dashed parallelograms) are observed at the base of the triangles. These extensions arise due to LET, as discussed in the text.  
\textbf{e},  Schematic representation of the negative bias triangles. Regions outlined by the solid back lines correspond to conventional transport, while those outlined by solid green lines (the tails) correspond to LET. Lengths corresponding to $E_\text{ST}$ are labeled. 
\textbf{f},  Transport details and energy level schematics for the electron triangle corresponding to the points denoted by the red square, purple diamond and teal cross in \textbf{d} and \textbf{e}. }
\end{figure*}

A simple rate model gives insight into when LET occurs. The rates in the model correspond to transitions between the states shown in Fig.~1b, and the corresponding lifetimes are the inverses of the rates. By calculating the expected amount of time required for an electron to pass through the system, we obtain a quantity proportional to the measured current $I$ (see Supplementary Information for the complete analysis). The slow rates are of interest here: the relaxation rate $\Gamma_\text{TS}$ from the triplet T(2,0) to the singlet S(2,0), the loading rate $\Gamma_\text{LS}$ of the singlet S(2,0) from the lead, and the unloading rate $\Gamma_\text{S}$ of the singlet S(2,0).  To focus on these rates and to develop intuition, we assume that all other rates are equal to a single rate, $\Gamma_\text{fast}$, an assumption that does not change the qualitative understanding.  The resulting proportionality for the current is
\begin{equation}
\nonumber
I\propto \frac{\Gamma_\text{fast}}{3+(\Gamma_\text{TS}+\Gamma_\text{LS})/\Gamma_\text{S}} .
\end{equation}
As this proportionality shows, the triplet tail is observed if and only if the sum of the triplet-singlet relaxation rate $\Gamma_\text{TS}$ and the loading rate from the lead $\Gamma_\text{LS}$ is not large compared to the escape rate $\Gamma_\text{S}$.  If the triplet-singlet relaxation rate is much faster than the escape rate, then the tail regime will be blockaded by electrons trapped in the S(2,0) state.  In our experiments, essentially no reduction in current ($\sim 5$\%) is observed moving from the bias triangle into the tail (from blue diamond toward teal cross in Fig.~3d).  Thus, electrons are rarely trapped in S(2,0), indicating that the triplet-singlet relaxation  rate $\Gamma_\text{TS}$ and the loading rate from the left lead $\Gamma_\text{LS}$ are both much less than $\Gamma_\text{S}$, itself a slow rate.  A similar calculation with the opposite bias shows that the condition for spin blockade is that the tripletÐsinglet relaxation rate in the (1,1) configuration is much slower than the fast rates, a far less stringent condition.

\begin{figure*}[t] 
\centering
\includegraphics[width=6.5in,keepaspectratio]{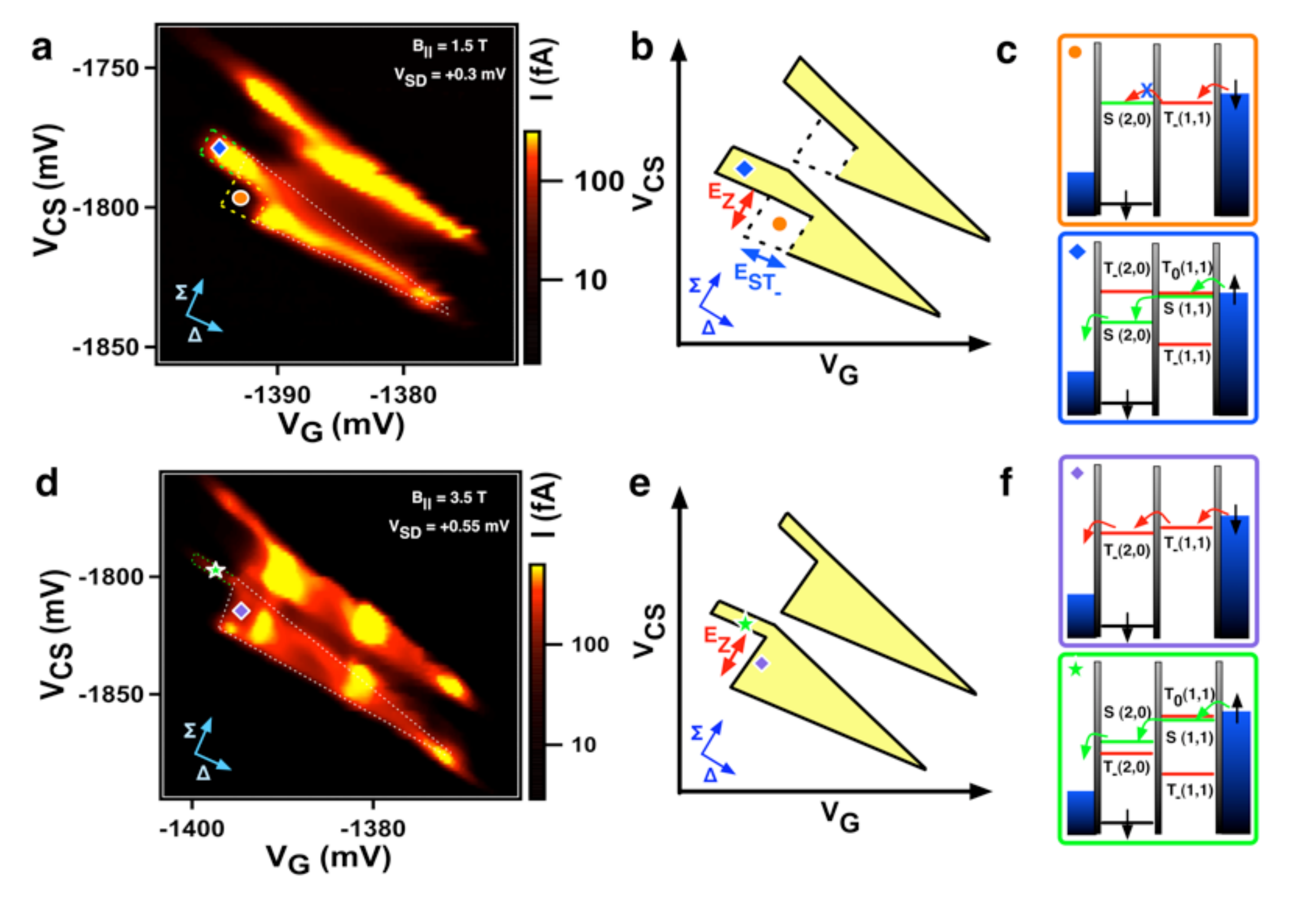}
\caption{
\textbf{Zeeman splitting, spin blockade and LET.  a},
Magnitude of the measured current with an in-plane magnetic field of 1.5~T applied such that $E_Z$ is less than $E_\text{ST}$. Spin blockade is reduced to a smaller region indicated by dashed yellow lines. Note that there is no conduction along the edges since the S(1,1) state is no longer energetically accessible when the (1,1) ground state T$_-$(1,1) aligns with the lead, preventing spin exchange.  In the region labeled with a blue diamond, a tail appears corresponding to transport through the excited spin singlet state. 
\textbf{b},  Schematic illustration of the bias triangles for small in-plane magnetic field. The energies $E_Z$ and $E_\text{ST-} = E_\text{ST}-E_Z$ can be extracted from the graph as indicated by red and blue arrows. The angles in this schematic are exaggerated relative to \textbf{a}.
\textbf{c},  Energy level schematics at the points labeled by the orange circle and blue diamond in \textbf{a} and \textbf{b}.
\textbf{d},  Magnitude of the measured current with an in-plane magnetic field of 3.5~T applied such that $E_Z$ is larger than $E_\text{ST}$. No blockade is observed inside the triangles. Excited states are observed inside the triangles as bright lines parallel to the triangle bases. The tail regions correspond to LET through excited spin singlet states.
\textbf{e},  Schematic illustration of the bias triangles for large in-plane magnetic field. The energy $E_Z$ can be extracted from the graph as indicated by the red arrow. The angles in this schematic are exaggerated relative to \textbf{d}.
\textbf{f},  Energy level schematics at the points labeled by the purple diamond and green star in \textbf{d} and \textbf{e}. }
\end{figure*}

To investigate the rapid tunneling between dots 1 and 2, and to understand the device physics, we have modeled the device numerically, as shown in Fig.~2c. Established methods are used to treat the various charge regions self-consistently, including trapped surface charge, ionized dopants, the two-dimensional electron gas (2DEG), and the device.\cite{r29}  The dopants are treated in the jellium approximation, while the inhomogeneous depletion of the 2DEG is treated semi-classically in a 2D Thomas-Fermi approximation.  For the gated region, a 2D Hartree-Fock basis of single-electron orbitals is obtained from the effective mass envelope equation, and a two-electron singlet wavefunction is constructed using the configuration interaction method, similar to ref.~\onlinecite{r14}  The results show that the bottom of the quantum dot confinement potential is nearly flat, with an oblong shape about 200~nm across.  General arguments suggest that the electron-electron interactions should dominate the kinetic energy in silicon for electrons separated by over 100~nm, causing two electrons to form a double dot.  The modeling results, shown in Fig.~2c, confirm that these general arguments give the correct intuition.  As is clear from the figure, the effective tunnel barrier between the two dots is low, consistent with lifetime-enhanced transport. We note that quantum dot splitting has been observed elsewhere, where it was attributed to deformation by a gate potential\cite{r24} or a local impurity.\cite{r25}  Although inhomogeneous confinement may also be present in our device, it is not needed to explain the double dot.

LET should be observable in many materials systems, provided the appropriate ratio of rates can be obtained.  Indeed, slow triplet-singlet relaxation and the preferential loading of triplets vs. singlets have both been observed in GaAs quantum dots, in pulsed-gate experiments.\cite{r30}  By analyzing the current vs. voltage data in our bias triangles (see Supplementary Information), we find that in the tail regime triplet loading occurs at a rate at least 1,000 times greater than singlet loading.  This ratio is 50 times greater than in previous observations of spin-dependent tunnelling,\cite{r30} which may lead to corresponding enhancements in spin readout.  The singlet loading is suppressed because its tunnel barrier to the external lead is larger than that the triplet state;\cite{r31} the relatively large effective mass of Si enhances this suppression.  The higher unloading rate is a consequence of the relatively small tunnel barrier between the two dots, as confirmed by numerical modeling.  Our bound on the singlet loading rate places a weak bound on triplet-singlet relaxation of $\Gamma_\text{TS} < 63,000\,\text{s}^{-1}$, although the actual value is expected to be much smaller.\cite{r32}

\begin{figure}[t] 
\centering
\includegraphics[width=1.8in,keepaspectratio]{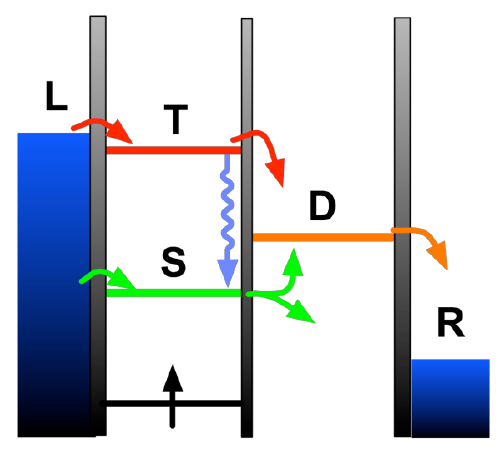}
\caption{
Energy levels and
associated tunneling rates for
transport through (2,0) and (1,1)
states of a double dot.}
\end{figure}

The phenomena described above of spin blockade and its complementary effect of LET can be unified by measurements of the system with an applied in-plane magnetic field.  In a magnetic field, the spin triplets are split linearly by the Zeeman energy ($E_Z = g\mu_\text{B}BS_Z$), where $\mu_\text{B}$ is the Bohr magneton, $S_Z$ is +1 for $|\uparrow \uparrow\rangle$, -1 for $|\downarrow \downarrow\rangle$ and 0 for $(|\uparrow \downarrow \rangle +|\downarrow \uparrow \rangle)/\sqrt{2}$. Since the $g$-factor is positive for silicon, the T$_-$ states shift lower in energy compared to the T$_0$ states, providing a technique for testing the interpretation of the data proposed above. 

Figure~4a-c (d-f) shows the energy diagrams for the cases where $E_Z$ is less (more) than $E_\text{ST}$. In Fig.~4a-c, the ground state of the (1,1) configuration is T$_-$(1,1), while that of the (2,0) configuration is S(2,0). Spin blockade now occurs in a smaller region than at $B = 0$, as indicated by dashed lines and the orange circle in Fig.~4a. Spin blockade is lifted in the conventional way when T$_-$(2,0) is lowered below T$_-$(1,1), corresponding to the triangular regions on the lower right in Fig.~4a,b.  Spin blockade is also lifted when the S(1,1) state can participate in transport (blue diamond).  However, the lifting of the blockade in this case is due to LET, because this S(1,1) state is an excited state of the (1,1) configuration, giving rise to a singlet tail in Fig.~4a,b.  This tail is a striking example of a generalization of LET: the singlet-triplet splitting is now inverted, and the LET is now due to long lifetimes in the singlet channel rather than the triplet channel. LET can be generalized to any situation where electron transport occurs through long-lived excited states, while lower energy states that would be metastably trapped are avoided.

\begin{figure}[t] 
\centering
\includegraphics[width=2.1in,keepaspectratio]{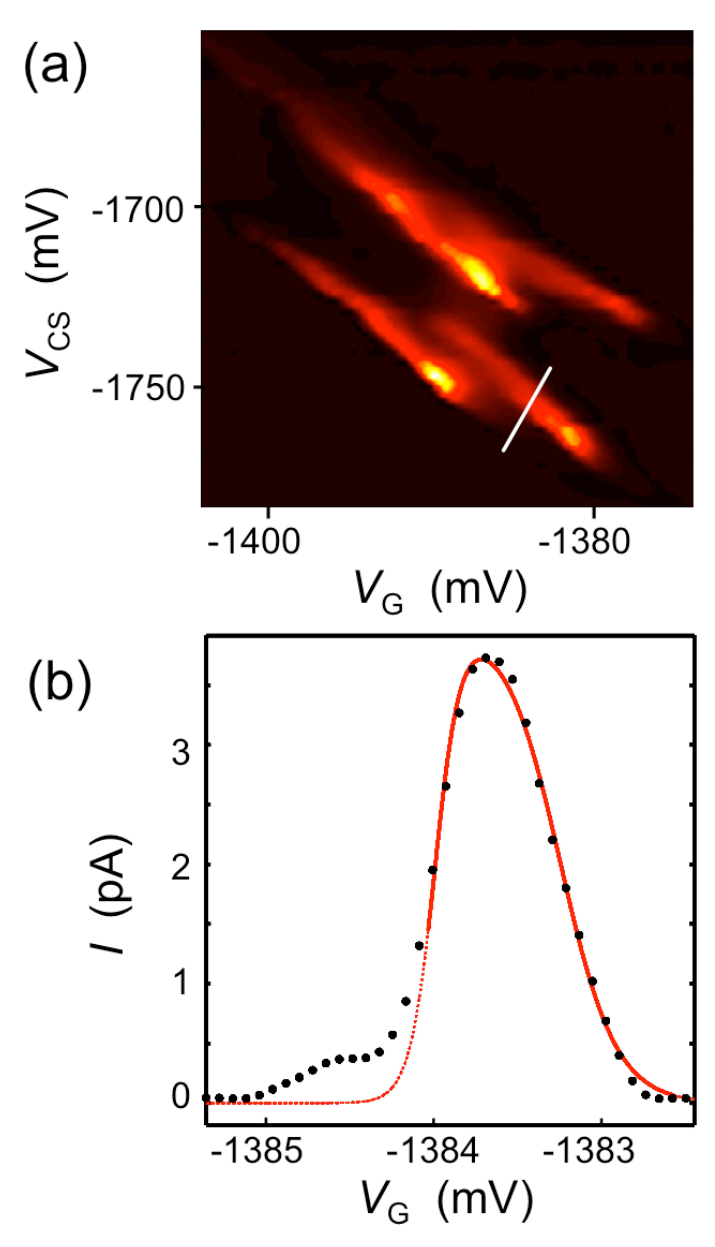}
\caption{
\textbf{Extraction of tunnel rates.  a},
The white line shows the
position of a typical data cut, taken
along a trajectory for which the
tunneling rate between singlet states
in the two dots is a constant. 
\textbf{b},  Current as a function of gate voltage
for the cut shown in \textbf{a}, plotted as
black circles. The data reveal a large
peak and a small peak that appears as
a shoulder on the left. For this cut,
the large peak corresponds to
tunneling via the triplet state, while
the small peak corresponds to
tunneling via the singlet state. The
data are fit to an elastic tunneling
model that takes the system through
the following transport cycle: 
$(1,0) \rightarrow (2,0) \rightarrow (1,1) \rightarrow (1,0)$. 
Tunneling processes that change the total
electron number involve tunneling
from the leads, while the (2,0) and
(1,1) states correspond to either
triplets or singlets. The solid line
shows the fit to the triplet peak, with
the extrapolation given by the faint,
dotted line. The singlet and triplet peaks were both fit in this way, over the entire bias
triangle, allowing us to extract the singlet loading and unloading rates throughout the
entire regime of Fig.~3. Moving parallel to the tail (perpendicular to the cut
shown in \textbf{a}), the loading of the singlet is approximately constant and equal to $1-2 \times
10^5$~s$^{-1}$. Also moving along the tail, the singlet unloading rate varies considerably, from as
large as $10^7$~s$^{-1}$ to as small as the loading rate itself, confirming the hypothesis that the tail
is observed due to unloading of the singlet state that is as fast or faster than its loading
rate. Results of our fitting also suggest that the height of the singlet peak (left side of 
\textbf{b}) is dominated by the S(2,0) escape rate $\Gamma_\text{S}$, 
providing a direct measurement of this quantity. }
\end{figure}

When $E_Z > E_\text{ST}$ (Fig.~4d-f), the ground state configurations are T$_-$(1,1) and T$_-$(2,0), and there should be no spin blockade because ground state transitions are allowed (purple diamond). Our measurements indeed show full triangles with no blockade. From the features visible inside the triangles (bright lines parallel to base) various excited states can be identified. LET also occurs through excited spin singlet states in this configuration, giving rise to a tail (green star). These data demonstrate the existence of long-lived electron spin states even in the presence of a finite magnetic field, a requirement for  various quantum operations.

\begin{acknowledgments}
We thank Nick Leaf for help with the data acquisition programs.  Support for this work was provided by NSA, LPS and ARO under contract number W911NF-04-1-0389, by the National Science Foundation through DMR-0325634 and DMR-0520527, and by DOE through DE-FG02-03ER46028.
\end{acknowledgments}
\vspace{.2in}
\appendix

\section*{SUPPLEMENTARY INFORMATION}
Appendix A contains an explanation of the rate model discussed in the main text. Appendix B
describes the extraction of loading and unloading rates in the double quantum dot.

\section{Discussion of the rate model.}
The energy level schematic shown in Fig.~5 illustrates electron transport through
a double quantum dot with one unpaired electron confined in the left dot at all times. The
labels in this figure have been simplified, for ease of labeling the rates. The incoming
electron from the left reservoir L can form either a spin singlet S(2,0), called S here, or a
spin triplet T(2,0), called T here. The energy levels corresponding to the spin states for
the (1,1) configuration are nearly degenerate and represented by D.

A rate model is used to gain insight into
when lifetime-enhanced transport (LET) occurs
through the double dot system. The states used in the
model correspond to the energy levels shown in the
figure, and the tunnel rates included are those
indicated by arrows in the diagram ($\Gamma_\text{LT}, \Gamma_\text{LS}, \Gamma_\text{TS}, \Gamma_\text{TD}, \Gamma_\text{SD}, \Gamma_\text{SR},$ and $\Gamma_\text{DR}$). The lifetimes used in the
calculations are the inverses of these rates, e.g., 
$\Gamma_\text{LT}=1/T_\text{LT}.$
We also define $\Gamma_\text{S}=\Gamma_\text{SD}+\Gamma_\text{SR}$, 
$\Gamma_\text{T}=\Gamma_\text{TS}+\Gamma_\text{TD}$, and
$\Gamma_\text{L}=\Gamma_\text{LT}+\Gamma_\text{LS}$, consistent with single
electron tunneling in the double quantum dot. The
probabilities associated with various electron
tunneling events are given by: 
$p_\text{LT}=\Gamma_\text{LT}/\Gamma_\text{L}$, 
$p_\text{LS}=\Gamma_\text{LS}/\Gamma_\text{L}$, 
$p_\text{TS}=\Gamma_\text{TS}/\Gamma_\text{T}$, 
$p_\text{TD}=\Gamma_\text{TD}/\Gamma_\text{T}$, 
$p_\text{SD}=\Gamma_\text{SD}/\Gamma_\text{S}$, and
$p_\text{SR}=\Gamma_\text{SR}/\Gamma_\text{S}$.

The model is Markovian, so the expected
time for exiting out of any state does not depend on the path to the state. Therefore, we
have:
\begin{eqnarray*}
T_\text{SR} &=& T_\text{S}+ p_\text{SD} T_\text{DR} , \\
T_\text{TR} &=& T_\text{T}+ p_\text{TS} T_\text{SR}+ p_\text{TD} T_\text{DR} , \\
T_\text{LR} &=& T_\text{L}+ p_\text{LT} T_\text{TR}+ p_\text{LS} T_\text{SR}  .
\end{eqnarray*}

By solving for the expected amount of time required for an electron to tunnel
through the system, $T_\text{LR}$, we obtain the quantity proportional to the measured current:
\begin{widetext}
\begin{equation*}
I\propto \frac{1}{T_\text{LR}}
=\frac{\Gamma_\text{DR}\Gamma_\text{L}\Gamma_\text{S}\Gamma_\text{T}}
{\Gamma_\text{LS}\Gamma_\text{SD}\Gamma_\text{T}+
\Gamma_\text{LT}\left( \Gamma_\text{SR}\Gamma_\text{TD}+\Gamma_\text{SD}\Gamma_\text{T})+
\Gamma_\text{DR}(\Gamma_\text{LT}(\Gamma_\text{S}+\Gamma_\text{TS})+
(\Gamma_\text{LS}+\Gamma_\text{S})\Gamma_\text{T}) \right)} .
\end{equation*}
\end{widetext}

Many of the rates are fast, especially the triplet channel. For developing intuition,
it is convenient to assume that all the fast rates are equal to a single rate $\Gamma_\text{fast}$ enabling us
to focus on the slow rates, which are the rates of interest: the loading of the singlet S(2,0),
$\Gamma_\text{TS}$ and $\Gamma_\text{LS}$, and the unloading of the (2,0) singlet $\Gamma_\text{S}$, resulting in (to lowest order),
\begin{equation}
\nonumber
I\propto \frac{\Gamma_\text{fast}}{3+(\Gamma_\text{TS}+\Gamma_\text{LS})/\Gamma_\text{S}} .
\end{equation}
Thus, LET is observed if and only if the sum of the triplet-singlet relaxation time and the
loading rate from the lead L is not large compared to the escape rate $\Gamma_\text{S}$, i.e., the second
term in the denominator is small. If the triplet-singlet relaxation rate is much faster than
the escape rate, then the tail regime will be blockaded. A similar calculation with the
opposite bias shows that the condition for spin blockade is that the triplet-singlet
relaxation rate is not large compared to the \textit{fast} rates, a far less stringent condition.

\section{Extracting loading and unloading rates for states in the double dot.}
Lifetime enhanced transport (LET) relies on the slow loading and fast unloading of a
low-lying energy state. In Fig.~3, this state is the (2,0) singlet state. A
quantitative analysis of our transport data in this figure allows us to compare loading and
unloading rates to confirm the LET hypothesis: that the unloading rate of the low-energy
(in this case, singlet) state is as fast or faster than its loading rate. We consider data
``cuts" through the lower bias triangle of Fig.~3d. Specifically, we
consider cuts running parallel to the base of the triangle, giving transport current as a
function of the gate voltage. The data exhibit behavior consistent with elastic tunneling
through a double dot,\cite{s1} with an additional inelastic component most likely due to nonresonant
tunneling between the two dots.\cite{s2} To model the elastic tunneling from the leads,
we consider a square tunnel barrier.\cite{s3} The modeled processes include tunneling from the
left lead (L) into the first dot (1), tunneling from the first dot into the second dot (2), and
tunneling from the second dot into the right lead (R). Our cut directions are chosen to
correspond to a constant ``detuning" between the energy levels in dots 1 and 2, so that for
a given cut the elastic tunneling rate from 1 to 2 is a constant. In this way, we can extract
quantitative values for the loading and unloading rates, obtaining a ratio $\Gamma_\text{unload}/\Gamma_\text{load}$, for
the (2,0) singlet state ranging from 1 at the tip of the triplet tail to 50-100 near the base of
the bias triangle. The rates themselves can be found in the figure caption. It is
impossible to separate the $\Gamma_\text{TS}$ triplet-singlet relaxation rate from the $\Gamma_\text{L}$ tunneling rate in
this model. However, $\Gamma_\text{load}$ provides an upper bound on both processes. Similarly, we can
obtain the current ratio $I_\text{T}/I_\text{S}$ for transport through the triplet vs. singlet channels,
obtaining a value of about 1000 along most of the triplet tail.. In Fig.~3d of the main text,
the singlet and triplet peaks are visibly separated, enabling tunability of the different
transport channels.

\end{document}